# Ion acoustic waves in the plasma with the power-law $q$-distribution in nonextensive statistics

Liu Liyan, Du Jiulin

*Department of Physics, School of Science, Tianjin University, Tianjin 300072, China*

**Abstract**: We investigate the dispersion relation and Landau damping of ion acoustic waves in the collisionless magnetic-field-free plasma if it is described by the nonextensive $q$-distributions of Tsallis statistics. We show that the increased numbers of superthermal particles and low velocity particles can explain the strengthened and weakened modes of Landau damping, respectively, with the $q$-distribution. When the ion temperature is equal to the electron temperature, the weakly damped waves are found to be the distributions with small values of $q$.

## 1. Introduction

The so-called ion acoustic waves are the low-frequency longitudinal plasma density oscillations. In the oscillations, electrons and ions are propagating in the phase space [1]. The ion acoustic waves were predicted first by Tonks and Langumir based on the fluid dynamics in 1929 [2]. The first experimental observation for the waves was reported in 1933 [3]. It is known that there are two models for the ion acoustic waves [4]: one is the continuum models, in which the plasma is treated as a fluid and so the fluid dynamics is used for its theoretical studies; the other one is based on the kinetic equations in statistical theory, where the distribution functions are used to describe the properties of the ion acoustic waves. As is well-known, Maxwellian distribution in Boltzmann-Gibbs (B-G) statistics is believed valid universally for the macroscopic ergodic equilibrium systems, but for the systems with the long-range interactions, such as plasma and gravitational systems, where the non-equilibrium stationary states exist, Maxwellian distribution might be inadequate for the description of the systems. For example, the experimentally measured phase velocity of the ion acoustic waves was 70% higher than the theoretical value derived under the presupposition that the plasma is described by Maxwellian distribution [5]. In the experiment for measuring the ion acoustic waves, the energy distribution of electrons may be actually not the Maxwellian one and hence we are hard to determine the valid electron temperature [6]. In fact, the non-Maxwellian velocity distributions for electrons in plasma were already measured in the experiment where the temperature gradient was steep [7]. And the non-Maxwellian velocity distributions for ions were also reported in the studies of the earth' plasma sheet, the solar wind, and elsewhere the long-range interacting systems containing plentiful superthermal particles, i.e., the particles with the speeds faster than the thermal speed [8, 9].

In recent years there has been an increasing focus on a new statistical approach, nonextensive statistics (or Tsallis statistics). It is described by a nonextensive parameter $q$. For $q \neq 1$, it gives power-law distribution functions and only when the parameter $q \to 1$ Maxwellian distribution is recovered [10]. It is thought to be a useful generalization of B-G statistics and to be appropriate for the statistical description of the long-rang interaction systems, characterizing the non-equilibrium stationary state [11]. Ref. [12] shown that, for electrostatic plane-wave propagation in plasmas, Tsallis formalism presents a good fit to the experimental data, while the standard Maxwellian distribution only provides a crude description. By restricting the value of $q$-parameter from the experimental data one can attain a good agreement between the theory and the experiment [13]. We will show here that the plasma described by $q$-distribution with $q < 1$ contains plentiful supply of superthermal particles. Furthermore, the flexibility provided by the nonextensive parameter $q$ enables one to obtain a good agreement between the theory and experiment.

The paper is organized as follows. In Sec. II, we study the dispersion relations and Landau damping in the new statistics with the power-law $q$-distributions. In Sec. III, the Landau damping is discussed under the different constraints on the values of $q$. Summary and conclusion are given in Sec. IV.

## 2. The generalized dispersion relations and Landau damping

In Maxwellian description in B-G statistics, the formulae of the dispersion relation and the ratio of Landau damping to the frequency of ion acoustic waves in the equilibrium plasma [1] are respectively

$$\frac{\omega_r^2}{k^2} = \frac{1}{1+k^2\lambda_{De}^2}\frac{K_B T_e}{m_i} + \frac{3}{2}v_{Ti}^2 \tag{1}$$

$$\frac{\gamma}{\omega_r} = -\sqrt{\frac{\pi}{8}}\left(\frac{1}{1+k^2\lambda_{De}^2}+3\frac{T_i}{T_e}\right)^{3/2}\left[\sqrt{\frac{m_e}{m_i}}+\left(\frac{T_e}{T_i}\right)^{3/2}\exp\left(-\frac{T_e/2T_i}{1+k^2\lambda_{De}^2}-\frac{3}{2}\right)\right] \tag{2}$$

where $\omega_r$ is the frequency of ion acoustic wave, $k$ is the wave number, $\lambda_{De}$ is the Debye length of electrons, $K_B$ is the Boltzmann constant, $\gamma$ is the Landau damping, $v_{Ti}$ is the thermal speed of ions. $T_e$ and $T_i$ are the temperature of electrons and ions, respectively, $m_i$ and $m_e$ are the mass of an ion and an electron, respectively. Here we will study the ion acoustic waves in the $q$-distribution descriptions in Tsallis statistics. Eq. (1) and (2) will be generalized in Tsallis statistics, which may describe their behaviors when the plasma is in the non-equilibrium stationary state.

First let us remind some basic facts about Tsallis statistics. In Tsallis statistics, the entropy has the form [10] of

$$S_q = K_B \frac{\int (f^q - f) d^3x d^3v}{1-q}, \tag{3}$$

where $f$ is the $q$-distributions, and the parameter $q$ different from unity specifies the degree of nonextensivity. The B-G entropy is recovered in the limit $q \to 1$. The basic property of Tsallis entropy is the nonadditivity or nonextensivity for $q \neq 1$. For example, for two systems A and B, the rule of composition [10] reads

$$S_q(A+B) = S_q(A) + S_q(B) + (1-q) S_q(A) S_q(B). \tag{4}$$

The $q$-equilibrium distribution function takes the power-law form. For one-dimensional case, it is given [14] by

$$f_0 = \frac{n_0 A_q}{\sqrt{\pi} v_T} \left[ 1 - (q-1) \frac{v^2}{v_T^2} \right]^{1/(q-1)} \tag{5}$$

$$A_q = \sqrt{1-q} \frac{\Gamma\left(\frac{1}{1-q}\right)}{\Gamma\left(\frac{1}{1-q} - \frac{1}{2}\right)} \quad 0 < q \leq 1, \text{ and } A_q = \frac{1+q}{2} \sqrt{q-1} \frac{\Gamma\left(\frac{1}{q-1} + \frac{1}{2}\right)}{\Gamma\left(\frac{1}{q-1}\right)} \quad q \geq 1,$$

where $n_0$ is the particle number density, $T$ is the temperature, $m$ is the mass and $v_T$ is the thermal speed, $v_T = \sqrt{2K_B T/m}$. It is worthy to notice that for $q > 1$ there is a thermal cutoff on the maximum value allowed for the velocity of the particles, namely $v < \sqrt{v_{T\alpha}/(q-1)}$. As expected, Maxwellian distribution is obtained in the limit $q \to 1$. Many classical questions in B-G statistics have been reconsidered in the framework of Tsallis statistics. Among them, let us write the generalized Boltzmann equation [15],

$$\frac{\partial f}{\partial t} + \mathbf{v} \cdot \nabla f + \frac{\mathbf{F}}{m} \cdot \nabla_v f = C_q(f), \tag{6}$$

where $C_q(f)$ is the $q$-collisional integral term, and $\mathbf{F}$ is the external force. It has been demonstrated that if $q > 0$ the time dependent solutions of the generalized equation (6) will evolve irreversibly towards the $q$-equilibrium distribution (5) and $C_q$ will vanish.

As we know, spacecraft measurements of plasma velocity distributions, both in the solar wind and in the planetary magnetospheres and magnetosheaths, have revealed that non-Maxwellian distributions are quite common. In many situations the distribution have a "suprathermal" power-law tail at high energies. This has been well modeled by the so-called κ-distribution [16], which, now we have known [17], is actually equivalent to the q-distribution Eq.(5). In other words, Eq.(5) and its leading results can be directly applicable to the above physical situations. In fact, in addition to the solar wind and the planetary plasma, the sun's interior plasma is the physical situation where Eq.(5) can be applicable to for the statistical description of it being the nonequilibrium stationary-state [18].

For a magnetic-field-free plasma which slightly departs from equilibrium, we use Eq.(6) and let the distribution function be $f_\alpha = f_{\alpha 0} + f_{\alpha 1}$, where the letter $\alpha$ in the subscript of $f$ denotes particle species ($\alpha=i, e$; $i$ for ion and $e$ for electron), $f_{\alpha 0}$ corresponds to the one-dimensional power-law $q$-equilibrium distribution (5), and $f_{\alpha 1}$ is the corresponding perturbation about the distribution (5). As one knows, the dynamical behavior of the plasma is governed by a combination of the generalized Boltzmann equation and Poisson equation [1]. Here we assume the plasma temperature to be high enough so that the $q$-collisional term in the generalized Boltzmann equation (6) is negligible. Neglecting high-order terms in the expansion of the distribution function and linearizing the equation, one finds

$$\frac{\partial f_{\alpha 1}}{\partial t} + \mathbf{v} \cdot \nabla f_{\alpha 1} + \frac{Q_\alpha}{m_\alpha} \mathbf{E_1} \cdot \nabla_v f_{\alpha 0} = 0 \tag{7}$$

and Poisson equation

$$\nabla \cdot \mathbf{E_1} = \frac{1}{\varepsilon_0} \sum_\alpha Q_\alpha \int f_{\alpha 1} d\mathbf{v} , \tag{8}$$

where $\mathbf{E_1}$ is the electric field produced by the perturbation and $Q_\alpha$ is charge of the particle.

We consider the direction of wave vector $\mathbf{k}$ to be along $x$-axis, and let $v_x = u$. Making Fourier transformation for $x$ and Laplace transformation for $t$ in Eq. (7), we have

$$i(ku - \omega) f_{\alpha 1} + \frac{Q_\alpha}{m_\alpha} E_1 \frac{\partial f_{\alpha 0}}{\partial u} = 0 . \tag{9}$$

Combined with Eq,(8), the following dispersion equation is obtained,

$$1 + \sum_\alpha \frac{\omega_{p\alpha}^2}{k^2} \int \frac{\partial \hat{f}_{\alpha 0}/\partial u}{\omega/k - u} du = 0 , \tag{10}$$

where $\omega_{p\alpha} = \sqrt{n_{\alpha 0} Q_\alpha^2 / \varepsilon_0 m_\alpha}$ is the naturally oscillating frequency of the plasma, $\hat{f}_{\alpha 0}$ is the normalized distribution function, $\hat{f}_{\alpha 0} = f_{\alpha 0}/n_{\alpha 0}$. Inserting the power-law distribution $f_{\alpha 0}$, Eq.(5), in Eq. (10), one readily gets

$$1 + \sum_\alpha \frac{1}{k^2 \lambda_{D\alpha}^2} \left[ \frac{1+q}{2} + \xi_\alpha Z_q(\xi_\alpha) \right] = 0 , \tag{11}$$

where $\lambda_{D\alpha} = v_{T\alpha}/\sqrt{2}\omega_{p\alpha}$ is the Debye length, $\xi_\alpha$ is a dimensionless parameter, defined as the ratio of the phase velocity, $v_\phi = \omega/k$, to the thermal speed $v_{T\alpha}$, namely, $\xi_\alpha = v_\phi/v_{T\alpha}$. $Z_q(\xi_\alpha)$ is the generalized plasma dispersion function in the context of Tsallis statistics,

$$Z_q(\xi_\alpha) = \frac{A_q}{\sqrt{\pi}} \int_{-\infty}^{+\infty} \frac{\left[1 - (q-1)x^2\right]^{2-q/(q-1)}}{x - \xi_\alpha} dx . \tag{12}$$

In the limit $q \to 1$, it is reduced to the standard form in B-G statistics [19],

$$Z(\xi_\alpha) = \frac{1}{\sqrt{\pi}} \int_{-\infty}^{+\infty} \frac{\exp(-x^2)}{x - \xi_\alpha} dx. \tag{13}$$

Obviously, there exists a singularity, $x = \xi_\alpha$, in the integrand of the dispersion function (12). Here we follow the line developed by Landau to deal with the singular point [20, 21]. The frequency is expressed in the complex form: $\omega = \omega_r + i\gamma_q$, where $\gamma_q$ is the generalized Landau damping that is related to $q$. For the weakly damped modes, the real part of dispersion function (12) is the Cauchy principal value, while the imaginary part is equal to half the residue of the integrand at the singularity, e.g.

$$Z_q(\xi_\alpha) = \frac{A_q}{\sqrt{\pi}} \mathrm{Pr} \int_{-\infty}^{+\infty} \frac{\left[1-(q-1)x^2\right]^{2-q/(q-1)}}{x - \xi_\alpha} dx + i A_q \sqrt{\pi} \left[1-(q-1)\xi_\alpha^2\right]^{2-q/(q-1)}. \tag{14}$$

Then Eq. (11) can be written as

$$1 + \sum_\alpha \frac{1}{k^2 \lambda_{D\alpha}^2} \left\{ \frac{1+q}{2} + \xi_\alpha \frac{A_q}{\sqrt{\pi}} \mathrm{Pr} \int_{-\infty}^{+\infty} \frac{\left[1-(q-1)x^2\right]^{2-q/(q-1)}}{x - \xi_\alpha} dx + i\xi_\alpha A_q \sqrt{\pi} \left[1-(q-1)\xi_\alpha^2\right]^{2-q/(q-1)} \right\} = 0. \tag{15}$$

In a plasma, if the electron temperature is much higher than the ion temperature, $T_e \gg T_i$, and the ion mass is much heavier than the electron's, $m_i \gg m_e$, the phase speed $v_\phi = \sqrt{K_B T_e / m_i}$ is much more than the thermal speed of ions $v_{Ti} = \sqrt{K_B T_i / m_i}$ but much less than that of electrons $v_{Te} = \sqrt{K_B T_e / m_e}$, $v_{Ti} \ll v_\phi \ll v_{Te}$. From the definition of the dimensionless parameter, $\xi_\alpha = v_\phi / v_{T\alpha}$, we get $\xi_e \ll 1$ and $\xi_i \gg 1$. Making series expansion for the integrand in Eq. (15) and integrating it, we obtain

$$1 + \frac{1}{k^2 \lambda_{De}^2} \frac{1+q}{2} - \frac{\omega_{pi}^2}{\omega^2}\left(1 + \frac{3}{(3q-1)} \frac{k^2 v_{Ti}^2}{\omega^2}\right) + \frac{i\sqrt{\pi} A_q}{k^2 \lambda_{De}^2} \xi_e \left[1-(q-1)\xi_e^2\right]^{2-q/q-1} + \frac{i\sqrt{\pi} A_q}{k^2 \lambda_{Di}^2} \xi_i \left[1-(q-1)\xi_i^2\right]^{2-q/q-1} = 0 \tag{16}$$

Inserting $\omega = \omega_r + i\gamma_q$ in Eq. (16) and making the real part of left side to be zero, we find

$$1 + \frac{1}{k^2 \lambda_{De}^2} \frac{1+q}{2} - \frac{\omega_{pi}^2}{\omega_r^2}\left[1 + \frac{3}{(3q-1)} \frac{k^2 v_{Ti}^2}{\omega_r^2}\right] = 0, \tag{17}$$

Thus we obtain the generalized dispersion relation,

$$\frac{\omega_r^2}{k^2} = \frac{1}{\frac{1+q}{2} + k^2 \lambda_{De}^2} \frac{K_B T_e}{m_i} + \frac{3}{3q-1} v_{Ti}^2. \tag{18}$$

As expected, in the limit $q \to 1$, Eq. (18) reduces to Eq. (1), the familiar result in B-G statistics. If the electron temperature is close to the ion temperature in the limiting case of long-wavelength $k\lambda_{De} \ll 1$, from Eq. (18), the generalized phase velocity becomes, $v_\phi = \omega_r/k = v_{Ti}\sqrt{2(3q+1)/(1+q)(3q-1)}$. Then the constraint $q > 1/3$ is imposed in order to keep the phase velocity positive. The traditional phase velocity in the framework of B-G statistics, $v_\phi = \sqrt{2} v_{Ti}$, is recovered in the limit $q \to 1$. It is easy to verify that the generalized phase velocity is faster than the traditional phase velocity when $q < 1$ and is slower than the traditional one when $q > 1$. If the electron temperature is much higher than the ion's, in the limit case of long-wavelength, the generalized phase velocity will be much higher than the ion thermal speed.

An interesting feature of ion acoustic waves in plasmas is that they should be damped even in the case of no collisions among particles [20], and the collisionless damping is explained as the interactions between the wave and particles moving with the speed close to the phase velocity [22]. Let the imaginary part of Eq. (16) to be zero, we obtain the generalized Landau damping as

$$\gamma_q = -\omega_r A_q \sqrt{\pi} \xi_i^3 \left\{ \frac{\lambda_{Di}^2}{\lambda_{De}^2} \frac{\xi_e}{\xi_i} \left[ 1 - (q-1)\xi_e^2 \right]^{2-q/(q-1)} + \left[ 1 - (q-1)\xi_i^2 \right]^{2-q/(q-1)} \right\}. \quad (19)$$

Using the dispersion relation (18), we find

$$\frac{\gamma_q}{\omega_r} = -\sqrt{\frac{\pi}{8}} A_q \left( \frac{1}{\frac{1+q}{2} + k^2 \lambda_{De}^2} + \frac{6}{3q-1} \frac{T_i}{T_e} \right)^{3/2} \left\{ \sqrt{\frac{m_e}{m_i}} + \left( \frac{T_e}{T_i} \right)^{3/2} \left[ 1 - (q-1) \left( \frac{T_e/2T_i}{\frac{1+q}{2} + k^2 \lambda_{De}^2} + \frac{3}{3q-1} \right) \right]^{2-q/(q-1)} \right\}. \quad (20)$$

In the limit $q \to 1$, Eq. (20) reduces to Eq. (2) in the framework of B-G statistics. In the above equation, the second term in the brace represents the contribution of the ions, which plays a main role, and the first term in the brace is related to the electrons, which is negligible for the fact that the electron mass is much lighter than the ion's.

## 3. The Landau damping and $q$-parameter

In this section, we will explore the extended Landau damping under different constraints on the value of $q$-parameter. Considering that Landau damping is related to the particle velocity distributions [1], we first investigate the properties of the power law $q$-distributions. In Fig. 1 (a), the curves of the distribution $\tilde{f} = \sqrt{\pi} v_{Ti} f / n_0 A_q$ as a function of the ratio of the particle velocity to the thermal speed with the constraint $q \le 1$ are plotted, and in Fig. 1 (b), the restriction is modified to $q \ge 1$. The values of the $q$-parameter in Fig. 1 (a) are 1 (dashed line), which is just the Maxwellian distribution, 0.6 (dotted line), and 0.2 (solid line). From Fig. 1 (a), we can see that in the case of the power law distributions, comparing with the Maxwellian distribution, there are more superthermal particles, i.e. particles with the speed faster than the thermal speed, when the value of $q$ is small. In Fig. 1 (b), the values of $q$ are 1.8 (dashed line), 1.4 (solid line), and 1.0 (dotted line). There is a cutoff in the tail when $q > 1$, namely, $v < \sqrt{v_{T\alpha}/(q-1)}$, and the maximum value allowed for the velocity of particles gets lower as $q$ increases. Thus we know the $q$-distributions with the constraint $q > 1$ are suitable for the description of systems containing a large number of low speed particles.

In Fig. 2, we plot the curves of $\gamma_q / (k v_{Ti})$ as a function of the ion dimensionless parameter $\xi_i = v_\phi / v_{Ti}$ with some selected values of the nonextensive parameter $q$. The selected values are 1.4 (dash-dotted line), 1.0 (dashed line), 0.6 (dotted line), and 0.4 (solid line). It shows that when the phase velocity is low or, equivalently, $\xi_i$ is small, the damping is heavier for higher values of $q$, but in the regions that the phase velocity being much faster than ion thermal speed, the waves are more damped for lower values of $q$. Now we discuss the reason for this phenomenon. As one knows Landau damping stems from the resonant interactions between the wave and particles moving with the speed close to the phase velocity, as a result, the damping conditions is determined by the number of resonant particles. On the other hand, we know that the number of the superthermal particles in the system obeying the power law distributions increases as the values of $q$ decrease as in Fig. 1 (a). Therefore when the phase velocity is much higher than the ion thermal speed there are more resonant superthermal ions in the systems with lower values of $q$. Consequently the interactions between the wave and ions are stronger and the waves are more damped for lower values of $q$. The power law distributions (5) with the constraint $q > 1$ describes the system composed of a large number of low velocity particles, hence heavy damping modes are formed in the region of low phase velocity.

The ratio of the damping to the frequency, $\gamma_q / \omega_r$, as a function of the ratio of the electron temperature to ion temperature in the limiting case of long-wavelength $k\lambda_{De} \ll 1$ is shown in Fig. 3. The selected values of $q$ are 1.4 (dash-dotted line), 1.0 (dashed line), 0.6 (dotted line), and 0.4 (solid line). It is obvious that for a fixed value of $q$ the damping is small when the electron temperature is much higher than ion temperature, and it reaches the maximum value when $T_e / T_i = 1$. First, we pay attention to the maximum damping presented in the case of equal electron and ion temperature. By Eq. (18), we can see that in such a case the generalized phase velocity of ion acoustic wave is low when $q > 1$, meanwhile for the case of the power law distributions with the same value of $q$, there are a large number of low velocity particles. As result, heavily damped modes appear. When $q < 1$, the generalized phase velocity is higher than the traditional phase velocity in B-G statistics and it increases as the value of $q$ decreases. From Fig. 1 (a), we see that the proportion of particles whose speed is more than two times of the thermal speed is small, so when the phase velocity is high, the number of particles moving with the speed close to the phase velocity is small. This will lead to the existence of weak damping modes. If the criterion used for a weak damping mode is that the imaginary part of $\omega$ shall be less than the real part divided by $2\pi$, i.e. $\gamma / \omega_r < 1/2\pi$, in Fig. 3 weak damping modes are seen when $q$ takes the values 0.4 and 0.6. As for the increased damping for smaller values of $q$ in the region of the electron temperature being much higher than ion temperature, it can be understood as a contribution by the increased number of superthermal particles in the tails close to the phase velocity.

## 4. Summary and Conclusion

In this paper we discuss the ion acoustic waves in plasmas with the power law

$q$-distributions in Tsallis statistics, and we obtain the extended dispersion relations and Landau damping. Unlike the description of Maxwellian distribution that most of the particles center around the thermal speed, the power law distributions characterize the systems containing plentiful supply of superthermal particles with the constraint $q<1$ or including a large number of low velocity particles by the restriction $q>1$. Thus the Landau damping which relies on the particle velocity distributions is related to the value of $q$-parameter. Weak damping modes are introduced for the power law distributions with small values of $q$ in the case of equal electron and ion temperature, while for Maxwellian system only heavily damped modes exist. When the phase velocity of ion acoustic wave is much faster than the ion thermal speed, such as in the case of the electron temperature being much higher than ion's, the extended Landau damping becomes stronger for lower values of $q$. Moreover, when the phase velocity is low, more heavily damped modes are found for larger values of $q$. All of these results are explained by the increased number of superthermal particles or low velocity particles contained in the plasma with the power law $q$-distributions.

Finally, it is should be emphasized that the physical state described by the $q$-distribution in Tsallis statistics is not the thermodynamic equilibrium. The nonextensive parameter $q$ was proved to relate to the temperature gradient and the potential energy of the system by the formula $K_B \nabla T + (1-q) Q_\alpha \nabla \varphi = 0$. Thus, the deviation of $q$ from unity qualifies the degree of the inhomogeneity of temperature or the deviation from the equilibrium [11]. Therefore, the properties of ion acoustic waves derived here are actually of the ones of plasmas in non-equilibrium stationary-state. Furthermore our results suggest that Tsallis statistics is suitable for the systems being the nonequilibrium stationary-state with inhomogeneous temperature and containing plentiful supply of the superthermal or low velocity particles.

## Acknowledgements

We would like to thank the National Natural Science Foundation of China under the grant No.10675088.

## References

[1] Li Ding and Chen Yinhua et al, Plasma Physics, Higher education press, Beijing, 2006；Xu Jialuan and Jin Shangxian, Plasma Physics, Nuclear Energy Press, Beijing, 1981.
[2] L. Tonks and I. Langmuir, Phy. Rev. **33**(1929)195.
[3] R.W. Revans, Phy. Rev. **44**(1933)798.
[4] Burton D. Fried and Roy W. Gould, The Phys. Fluids **4**(1961)139.
[5] A.Y. Wong, N.D' Angelo, and R. W. Motley, Phy. Rev. Lett. **9**(1962)415.
[6] I. Alexeff and R.V. Neidigh, Phy. Rev. **129**(1963)516.
[7] E. T. Sarris, S. M. Krimigis, A. T. Y. Lui, K. L. Ackerson, L. A. Frank, and D. J. Williams, Geophys. Res. Lett. **8**(1981)349. D. J. Williams, D. G. Mitchell, and S. P.

   Christon, Geophys. Res. Lett. **15**(1988)303.
[8] J.T. Gosling, J.R. Asbridge, S.J. Bame, W.C. Feldman, R.D. Zwickl, G. Paschmann, N. Sckopke, and R. J. Hynds, J. Geophys. Res. [Space Phys.] **86**(1981)547.
[9] J.M. Liu, J.S. De Groot, J.P. Matte, T.W. Johnston, R.P. Drake, Phys. Rev. Lett. **72** (1994)2717.
[10] C. Tsallis, J. Stat. Phys. **52**(1988)479.
[11] J.L. Du, Europhys.Lett. **67**(2004)893; J.L. Du, Phy. Lett. A **329**(2004)262.
[12] J.A. S. Lima, R. Silva, J. Santos, Phys. Rev. E **61**(2000)3260.
[13] R. Silva, J.S. Alcaniz, J.A. S. Lima, Physica A **356**(2005)509.
[14] R. Silva, Jr., A. R. Plastino, and J. A. S. Lima, Phys. Lett. A **249**(1998)401.
[15] J.A.S. Lima, R. Silva, A.R. Plastino, Phys. Rev. Lett. **86**(2001)2938.
[16] S.R.Cranmer, Astrophys. J. **508**(1998)925.
[17] M.P.Leubner, Astrophys. Space Sci. **282**(2002)573; Astrophys. J. **604**(2004)469; L N Guo, J L Du and Z P Liu, Phys. Lett. A **367**(2007)431.
[18] J.L. Du, Europhys.Lett. **75**(2006)861.
[19] B.D. Fried, M. Gell-Mann, J.D. Jackson, and H.W. Wyld, J. Nuclear Energy: Part C 1, 190，1960.
[20] L. D. Landau, J. Phys. USSR **10**(1946 )25.
[21] N.A. Krall and A.W. Trivelpiece, Principle of Plasma Physics, McGraw-Hill, Kogakusha, 1973.
[22] D. Bohm, E. Gross, Phys. Rev. 75 (1949) 1851; D. Bohm, E. Gross, Phys. Rev. **75**(1949)1864.